\documentclass[3p,times,twocolumn]{elsarticle}
\pdfoutput=1

\usepackage[english]{babel}

\usepackage{amssymb}

\usepackage[figuresright]{rotating}

\journal{arXiv}

\begin{document}

\begin{frontmatter}




\title{Precision Electroweak Measurements at FCC-ee}

\tnotetext[t1]{Proceedings of the 37th International Conference on High Energy Physics - Valencia (Spain) - 2-9 July 2014}


\author{Roberto Tenchini}

\address{INFN Sezione di Pisa - Largo Bruno Pontecorvo, 3 - 56127 Pisa - Italy}

\begin{abstract}
The prospects for electroweak precision measurements at the Future Circular Collider with electron-positron beams (FCC-ee) are discussed. The Z mass and width, as well as the value of the electroweak mixing angle, can be measured with very high precision at the Z pole thanks to an instantaneous luminosity five to six order of magnitudes larger than LEP.  At centre-of-mass energies around 160 GeV, corresponding to the WW production threshold, the W mass can be determined very precisely with high-statistics cross section measurements at several energy points. Similarly, a very precise determination of the top mass can be provided by an energy scan at the $\mathrm{t \bar t}$ production threshold, around 350 GeV.
\end{abstract}


\end{frontmatter}


\section{Introduction}
\label{Intro}

The Future Circular Collider (FCC) concept builds on the successful experience with the LEP-LHC tunnel. A possible long-term strategy for high-energy physics at colliders, after the exploitation of the LHC following its High Luminosity upgrade, is based on a tunnel of about 100 km circumference, which takes advantage of the present CERN accelerator complex. In a first phase the tunnel could host an $\mathrm{e^+e^-}$ collider (FCC-ee) at centre-of-mass energy between the Z pole and above the $\mathrm{t \bar t}$ production threshold, with a subsequent proton proton collider as an ultimate goal. Together with a possible electron-proton option, the project would provide 50 years of physics at the highest energies.

\begin{figure}[t]  
\begin{center}
    \includegraphics[width=0.8\linewidth]{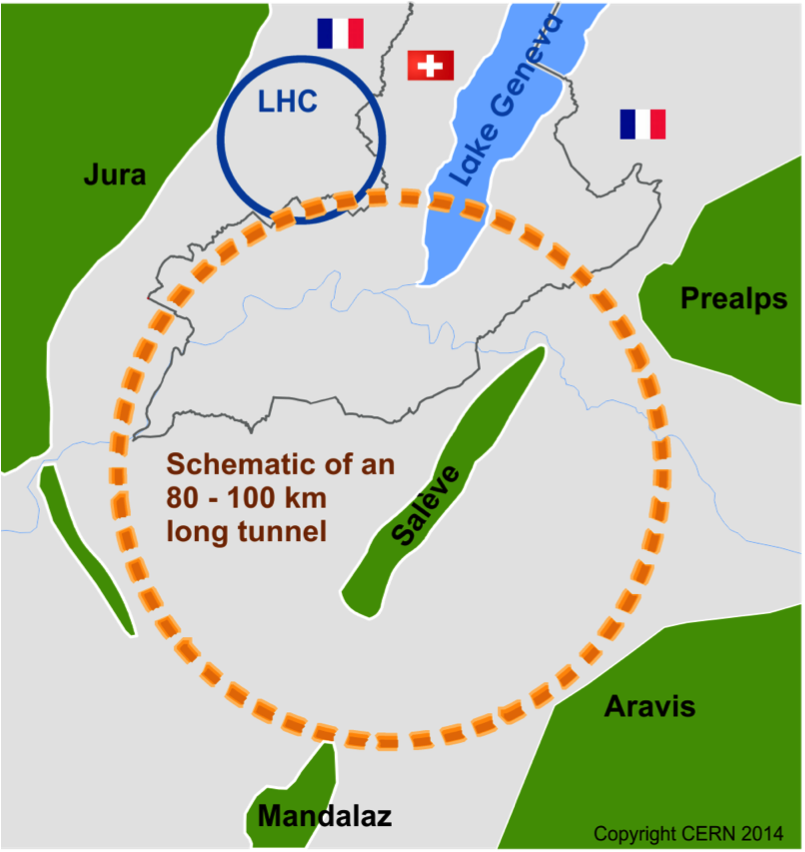} 
\caption{A possible location of the Future Circular Collider in the Geneva area.}
\label {fig:FFC-location}
\end{center}
\end{figure}

Future Circular Collider study groups have been formed, aiming at a Conceptual Design Report and a review cost in time for next European Strategy milestone (2018-2019). The groups should define the infrastructure requirements and the physics potential of the various steps. As can be seen from Fig.~\ref{fig:FFC-location}, a 80 to 100 km tunnel can fit in the Geneva area. A similar project, albeit with smaller size, is being considered in China (CEPC).

The FCC-ee will provide very high luminosity for a broad physics program~\cite{Gomez-Ceballos:2013zzn}. Here we discuss the physics potential for electroweak measurements with

\begin{itemize}
\item beams of 45.6 GeV (Z pole), 
\item beams in the range 80 - 100 GeV (at and just above the WW production threshold),
\item beams in the range 170 - 180 GeV (at and just above $\mathrm{t \bar t}$ production threshold), with the potential of reaching 250 GeV.
\end{itemize}

The prospects for FCC-ee as an Higgs factory, with beams of 120 GeV, are discussed elsewhere in these proceedings~\cite{Manqi}.

\section{The FCC-ee accelerator}

The potential of a very large $\mathrm{e^+e^-}$ circular collider has drawn considerable attention following the Higgs boson discovery.
The Higgs boson mass is low and not far from the hedge of LEP sensitivity,  about 115 GeV, which is only 10\% lower than the mass actually measured at LHC, i.e. 125 GeV. The highest centre-of-mass energy attainable at a circular collider is limited by radiation losses by the beam. 
Synchrotron energy loss per turn goes as $E^4/r $, where $E$ is the beam energy and $r$ the radius of the ring. An increase of the radius by a factor three, which is roughly the ratio between the proposed FCC and the LEP radii,  would be sufficient to produce the 125 GeV Higgs boson with about one half of the RF power utilised at LEP . A collider with such a large ring would also give an excellent opportunity to perform precision measurements at the Z pole, at the WW production threshold and at the $\mathrm{t \bar t}$ production threshold, providing a unique discovery potential by combining precision measurements and direct searches (e.g. search for right-handed neutrinos~\cite{Alain}).

The present conceptual design of the machine is driven by the requirement of precision Higgs physics at a centre-of-mass energy of 240 GeV and foresees a power consumption a factor five higher than LEP, still manageable, with a multi-bunch scheme. The same total power can be exploited to increase the centre-of-mass energy, with a reduction of the number of bunches and, consequently, of the luminosity. This way the $\mathrm{t \bar t}$ production threshold can be reached, opening a window to top physics and in particular to a precise measurement of the top mass.
On the other hand, the centre-of-mass energy can be lowered and the RF power used to increase considerably the number of bunches and, therefore, the luminosity. Recent studies~\cite{crab} have shown that at the Z one can introduce "crab waist" collision techniques and increase the luminosity even further.

\vspace{0.25cm}

\vspace{0.25cm}
\begin{figure}[t]  
\begin{center}
    \includegraphics[width=1.0\linewidth]{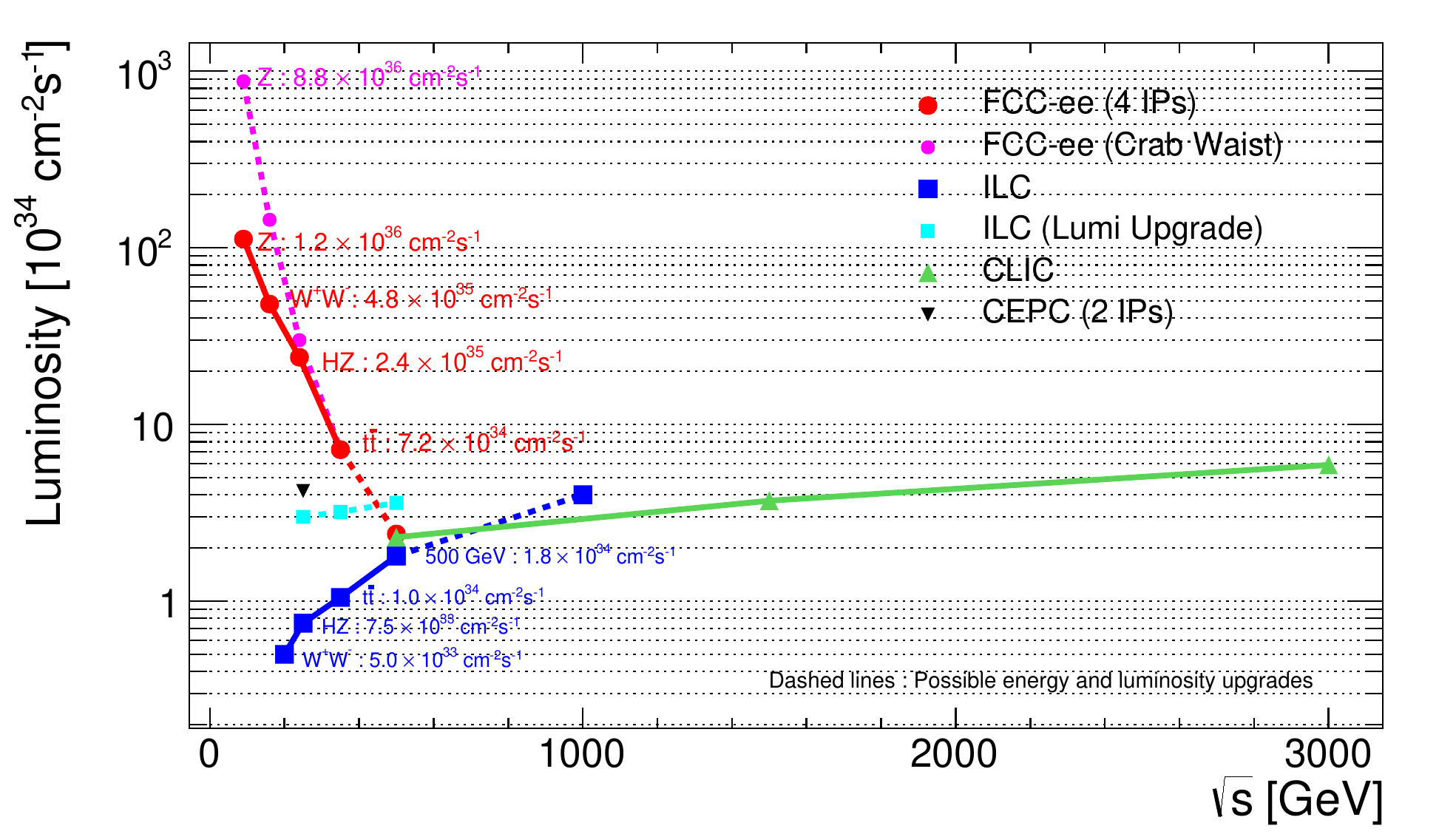} 
\caption{Instantaneous luminosity, in units of $10^{34} \mathrm{cm^{-2} s^{-1}}$, expected at FCC-ee (full red line), in a configuration with four interaction points operating simultaneously, as a function of the centre-of-mass energy. For illustration, the luminosities expected at linear colliders, ILC (blue line) and CLIC (green line), are indicated in the same graph. The plot includes further luminosity and energy upgrades for ILC and FCC-ee (dashed lines), under discussion at the time of writing these proceedings.}
\label {fig:luminosity}
\end{center}
\end{figure}

The FCC-ee design is based on an accelerator ring with a storage ring delivering continuous top-up injection. The storage ring compensates for the small beam lifetime caused by Bhabha scattering and loss of particles in collisions, providing a constant level of luminosity~\cite{Blondel:2011fua}. The multi bunch operation foresees more than16000 bunches with beams of 45.6 GeV (Z Pole) and about 100 bunches with beams of 175 GeV  ($\mathrm{t \bar t}$ production threshold). Figure~\ref{fig:luminosity} shows the FCC-ee expected instantaneous luminosity as a function of the centre-of-mass energy, if the luminosity is delivered at four interaction points. The highest luminosity is reached at the Z pole, as expected from the previous considerations. The behaviour is clearly complementary to linear colliders: much higher luminosity can be reached at a centre-of-mass energy well above 350 GeV, while linear colliders can potentially reach a much higher centre-of-mass energy.

\section{Physics at the Z pole}

A unique feature of an electron positron circular collider is the possibility to perform a high precision measurement of one of the pillars of electroweak fits: the Z mass trough the line-shape scan. Here the key point is the knowledge of the centre-of-mass energy during the scan, which is by far the dominating systematic uncertainty in the Z mass measurement. At LEP a precision of $2 \times 10^{-5}$ was reached~\cite{Z-Pole} with the technique of resonant depolarisation~\cite{Arnaudon:1994zq}. At FCC, as explained below, the precision could be increased by at least one order of magnitude: this method would lead to an uncertainty of 100 keV on both the Z mass and width. 

In a circular collider, transverse polarisation builds up naturally because of the Solokov-Ternov effect~\cite{Ternov}, which is related to bremsstrahlung. A magnet providing horizontal B field, positioned along the ring,  can be used to tilt the electron spin, inducing Thomas precession. If the electron spin turns out to be in phase at next turn, the horizontal B field will eventually rotate the spin in the accelerator plane, destroying polarisation. (Polarisation in the horizontal plane is affected by depolarising effects.) The {\it on phase} condition can be used to determine accurately  the beam energy. The beam polarisation is measured by means of a laser with Compton scattering, which depends on the electron spin direction. At LEP uncertainties on the beam energy of the order of 100 KeV were achieved in dedicated polarisation runs, however the final systematic uncertainty was  much larger because the polarisation had to be transported to collision physics runs. In the extrapolation several effects had to take into account such as the impact of earth tides, of geological shifts (e.g. the water level in the Geneva lake), temperature variations, parasitic currents (e.g. railways trains), etc. At FCC it is possible to dedicate a few of the many bunches to this purpose, and monitor the beam energy continuously, without requiring an extrapolation, with the goal of being affected by statistical uncertainties only. 

Polarisation is very interesting also for another reason: spin rotators can be employed to achieve longitudinal polarisation. Longitudinal polarisation is the key to sort out the $\sin^2 \theta_W$ puzzle, which is the long standing difference between the determination of the electroweak mixing angle from the left-right asymmetry, $\mathrm{A_{LR}}$, and the $\mathrm{Z \to b \bar b}$ forward-backward asymmetry~\cite{Z-Pole}. With two polarised beams a measurement of $\mathrm{A_{LR}}$ can be performed without an independent knowledge of beam polarisation by alternating the polarisation of the two beams~\cite{Blondel:1987wr}. In addition the b electroweak couplings can be determined, by means of a measurement of the polarised forward-backward  $\mathrm{Z \to b \bar b}$ asymmetry and of the  $\mathrm{Z \to b \bar b}$ partial width.

A very high statistic run at the Z pole is also very interesting to increase the precision of other observables, such as the ratio between the leptonic and hadronic Z decay widths, leading to improved determination of $\alpha_{s}$. A precise test of lepton universality can also be performed by comparing Z decays to electron, muon and tau pairs.

Another important subject concerns the number of quark-lepton generations. At LEP the number of neutrino families was determined from the total hadronic cross section at the Z peak and the final uncertainty was dominated by the theoretical knowledge of the Bhabha cross section used for the luminosity normalisation. Theory calculations need to be improved beyond the current $10^{-3}$ level to gain precision with this method. Potentially more promising is the use of Z radiative returns by running  at higher centre-of-mass energy, i.e. the process $\mathrm{e^+e^- \to Z \gamma}$, leading to the detection of a monochromatic photon for invisible Z decays. This second method was limited by statistics at LEP, but would give a very precise determination of the Z invisible width at FCC-ee. Radiative Z returns could also be used to perform an extensive search for sterile neutrinos~\cite{Alain}.

\section{Physics at the WW production threshold}

With the integrated luminosity expected at FCC-ee in one or two years of running at a centre-of-mass energies between 160 and 200 GeV one can expect  $O (10^8)$ WW pairs, four orders of magnitudes more than LEP~\cite{Schael:2013ita}.  A run at 160 GeV is particularly interesting, since the WW production cross section is very sensitive on the value of the W mass. Again, to perform a precise measurement of the mass, an accurate determination of the centre-of-mass energy is required.

At high energy the resonant depolarisation technique is getting less and less usable to determine the centre-of-mass energy because of the growing beam energy spread of the machine, which lowers the degree of beam polarisation. At LEP transverse polarisation was achieved only up to a beam energy of about 60 GeV. However, the beam energy spread is expected to be lower at FCC than LEP because of the larger radius of the accelerator and, based on LEP experience, beam polarisation should be available at the WW threshold, with the potential of achieving a very precise measurement of the W mass. This possibility will allow to take full advantage of the increase in WW statistics of four orders of magnitudes with respect to LEP and, extrapolating from LEP results, to reach a precision of 0.5 MeV with the threshold scan method. 

One can also expect improvements for the direct W mass reconstruction method based on di-jet invariant mass, which reached a precision of about 30 MeV at LEP~\cite{Schael:2013ita}. The systematic uncertainties in the semi-leptonic channel, which is not affected by colour-reconnection effects, are dominated by the understanding of the jet energy scale and jet angular resolution, in the range 10 - 20 MeV for the final LEP measurements. Very high statistics jet studies, better detectors and improved Monte Carlo simulations can lower considerably these uncertainties. From the di-jet invariant mass a precise direct determination of the W width can be obtained, too. 

At around 200 GeV centre-of-mass energy, there is also excellent potential for very precise measurements of boson self-couplings. Here the key is the possibility to measure differential distributions with very high statistics. As an example, for the charged triple gauge couplings, the $\mathrm{W^-}$ production angle in $\mathrm{e^+ e^- \to W^+ W^-}$ and the four W bosons decay angles can be used to separate the various polarisation components and compare their intensity with the predictions of the standard model. Deviations from the prediction are expected for anomalous gauge couplings and can be used to investigate the gauge structure of new physics. Precisions at the level of per mille or better for charged couplings can be expected.

\vspace{0.25cm}
\begin{figure}[t]  
\begin{center}
    \includegraphics[width=0.8\linewidth]{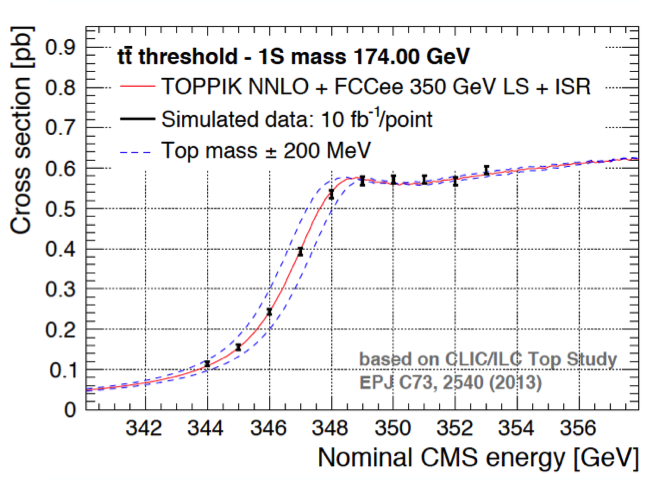} 
\caption{Simulation of a scan around the  $\mathrm{t \bar t}$ threshold at FCC-ee~\cite{Simon,Seidel:2013sqa}. Each centre-of-mass energy point corresponds to 10~$\mathrm{fb^{-1}}$, about one week of running at FCC-ee with the expected instantaneous luminosity.  The solid curve is taken from a NNLO calculation~\cite{Hoang:1999zc, Hoang:1998xf}, where QED ISR and FCC-ee beam parameters have been included. It assumes a top mass of 174 GeV in the 1S mass scheme, which is shifted by 200 MeV in opposite directions in the two dashed curves. }
\label {fig:topscan}
\end{center}
\end{figure}

\section{Physics at the $\mathrm{t \bar t}$  production threshold}

The FCC-ee can also be a very clean top factory with over one million $\mathrm{t \bar t}$ pairs produced in a few years in a very clean experimental environment. The most important achievement in this area would be a sizeable jump in precision in the determination of the top mass, which is currently determined at hadron colliders with an experimental precision of about 0.5\%. The theoretical interpretation of the experimental measurements with the current precision is already challenging, essentially because the top quark colour charge is transferred to the b quark in the $\mathrm{t \to Wb}$ decay, yielding a set of final state hadrons that cannot unambiguously be related to a single initial top quark. The measurement of the top mass parameter from the invariant mass of decay products is therefore affected by an uncertainty of the order of $\mathrm{\Lambda_{QCD} \sim 500~MeV}$, which is inherent in the method itself. 

The measurement of the inclusive cross section for the process $\mathrm{e^+ e^- \to t \bar t}$, which is very sensitive to the top mass at top pair production threshold, offers the opportunity to avoid the above mentioned problem because of the colour singlet nature of the production mechanism. The use of this process for a well-defined measurement of the top mass parameter has been already stressed many times in the context of of ILC studies~\cite{Simon,Seidel:2013sqa,Hoang:1999zc, Hoang:1998xf}.

At FCC-ee one could perform a top mass measurement with essentially the same precision of a linear collider, for the same integrated luminosity. A circular collider, and in particular the FCC-ee,  has the advantage of a low beamstrahlung level, with a narrow beam energy spectrum having a very low tail. As a consequence the cross section just above threshold is essentially independent on the top mass, which is a useful feature for combined fits of the top mass and other parameters (e.g. the top width, $\alpha_s$ or the top-Higgs Yukawa coupling). This expected behaviour has been recently checked with tools developed in the context of ILC studies, where typical FCC-ee beam parameters have been included~\cite{Simon,Seidel:2013sqa}, as shown in Fig.~\ref{fig:topscan}.

If a statistics of one million $\mathrm{t \bar t}$ pair is considered, with an accurate knowledge of the beam energy spectrum and an independent precise measurement of $\alpha_s$ one could aim to a statistical precision of 10 MeV on the top mass. The final precision will eventually depend on the theoretical uncertainty related to higher order QCD corrections.

The study of top decays produced in  $\mathrm{e^+ e^-}$ collisions is also very promising for the search of rare processes, such as decays induced by Flavour Changing Neutral Currents~\cite{Khanpour:2014xla}. Running well above the $\mathrm{t \bar t}$  production threshold  at FCC-ee would also be possible, a centre-of-mass energy of 500 GeV would require tripling the RF power, with the possibility of studying the associated production of top pairs and vector/scalar bosons.

\vspace{0.25cm}
\begin{figure}[t]  
\begin{center}
    \includegraphics[width=1.1\linewidth]{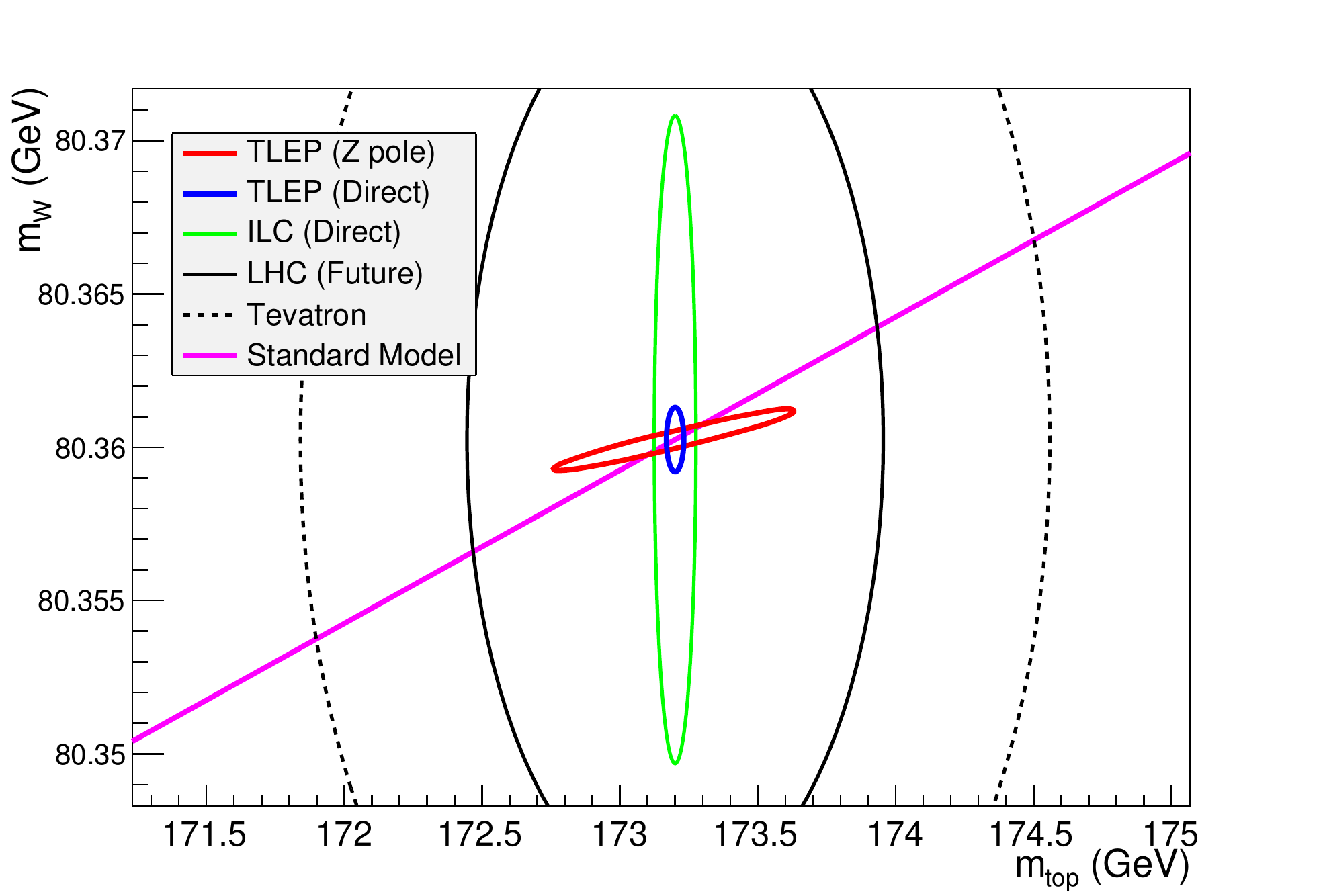} 
\caption{The 68 \% C.L. contour in the $\mathrm{m_{top}}$ -  $\mathrm{m_{W}}$ plane  as expected for FCC-ee (indicated as TLEP in the figure) and other accelerators. The blue line indicates the expected contour from direct W and top mass measurements, while the red line gives the expected precision from a fit of Z-pole observables.}
\label {fig:W-top-plane}
\end{center}
\end{figure}

\section{Summary of the goals and final remarks}

In conclusion a very large circular collider with electron-positron beams, taking data at centre-of-mass energies from 90 to 350 GeV and above could provide an impressive jump in precision in the knowledge of most electroweak observables. Potentially it could be used to measure the Z mass and width with an uncertainty of 100 KeV, the W mass with an uncertainty of 500 KeV, the top mass with an uncertainty of 10 MeV. The real achievable precision must be evaluated with studies in realistic conditions, and after having reviewed in details the many experimental and theoretical challenges.

If the above goals are achieved, the contour line in the celebrated $\mathrm{m_{top}}$ vs  $\mathrm{m_{W}}$ plot could evolve from today to FCC-ee, passing through a series of successive steps in precision at various accelerators, as shown in Fig.~\ref{fig:W-top-plane}.
Electroweak fits have historically anticipated the top and Higgs boson mass values, which such a jump in precision they could provide discriminating power for new physics. Sensitivity to new physics energy scales up to 100 TeV can be envisioned~\cite{Ellis}.

Nowadays particle physics is not facing a standard model, but rather an exceptional model, which must be challenged at very high precision.
While the hope is to detect soon signals of new physics, it is unlikely that a full scenario will appear at short-medium term.
At this point in time a precision $\mathrm{e^+ e^-}$ machine looks like a natural and effective next step. 
A circular collider would provide very high precision measurements of electroweak observables  (including a very detailed study of the Higgs boson) and it would pave the way to next high energy hadron collider.

\section*{Acknowledgements}

I would like to thank Patrizia Azzi, Alain Blondel, Patrick Janot and Frank Simon for their help in preparing the talk and for providing essential information.




\nocite{*}
\bibliographystyle{elsarticle-num}
\bibliography{tenchini-ICHEP2014}

\begin{thebibliography}{10}
\expandafter\ifx\csname url\endcsname\relax
  \def\url#1{\texttt{#1}}\fi
\expandafter\ifx\csname urlprefix\endcsname\relax\def\urlprefix{URL }\fi
\expandafter\ifx\csname href\endcsname\relax
  \def\href#1#2{#2} \def\path#1{#1}\fi

\bibitem{Gomez-Ceballos:2013zzn}
M.~Bicer, et~al., {First Look at the Physics Case of TLEP}, JHEP 1401 (2014)
  164.
\newblock \href {http://arxiv.org/abs/1308.6176} {\path{arXiv:1308.6176}},
  \href {http://dx.doi.org/10.1007/JHEP01(2014)164}
  {\path{doi:10.1007/JHEP01(2014)164}}.

\bibitem{Manqi}
M. Ruan, Higgs Physics at the FCC-ee, these proceedings.

\bibitem{Alain}
A. Blondel, Heavy neutrino hunting in Higgs- and Z decays at high luminosity
  Higgs and Z factory, these proceedings.

\bibitem{crab}
D. Shatilov and A. Bogomyagkov, Presented at the 7th TLEP-FCCee workshop, 19-21
  June 2014, CERN, URL http://indico.cern.ch/event/313708/.

\bibitem{Blondel:2011fua}
A.~Blondel, F.~Zimmermann, {A High Luminosity $e^+ e^-$ Collider in the LHC
  tunnel to study the Higgs Boson }\href {http://arxiv.org/abs/1112.2518}
  {\path{arXiv:1112.2518}}.

\bibitem{Z-Pole}
{The ALEPH, DELPHI, L3, OPAL, SLD Collaborations, the LEP Electroweak Working
  Group, the SLD Electroweak and Heavy Flavour Groups}, {Precision Electroweak
  Measurements on the Z Resonance}, Phys. Rept. 427 (2006) 257--454.
\newblock \href {http://arxiv.org/abs/hep-ex/0509008}
  {\path{arXiv:hep-ex/0509008}}, \href
  {http://dx.doi.org/10.1016/j.physrep.2005.12.006}
  {\path{doi:10.1016/j.physrep.2005.12.006}}.

\bibitem{Arnaudon:1994zq}
L.~Arnaudon, R.~Assmann, A.~Blondel, B.~Dehning, P.~Grosse-Wiesmann, et~al.,
  {Accurate determination of the LEP beam energy by resonant depolarization},
  Z.Phys. C66 (1995) 45--62.
\newblock \href {http://dx.doi.org/10.1007/BF01496579}
  {\path{doi:10.1007/BF01496579}}.

\bibitem{Ternov}
A.~Sokolov, I.~Ternov, {On Polarization and spin effects in the theory of
  synchrotron radiation}, Sov.Phys.Dokl. 8 (1964) 1203--1205.

\bibitem{Blondel:1987wr}
A.~Blondel, {A Scheme to Measure the Polarization Asymmetry at the $Z$ Pole in
  {LEP}}, Phys.Lett. B202 (1988) 145.
\newblock \href {http://dx.doi.org/10.1016/0370-2693(88)90869-6}
  {\path{doi:10.1016/0370-2693(88)90869-6}}.

\bibitem{Schael:2013ita}
{The ALEPH, DELPHI, L3, OPAL Collaborations and the LEP Electroweak Working
  Group}, {Electroweak Measurements in Electron-Positron Collisions at
  W-Boson-Pair Energies at LEP}, Phys.Rept. 532 (2013) 119--244.
\newblock \href {http://arxiv.org/abs/1302.3415} {\path{arXiv:1302.3415}},
  \href {http://dx.doi.org/10.1016/j.physrep.2013.07.004}
  {\path{doi:10.1016/j.physrep.2013.07.004}}.

\bibitem{Simon}
F. Simon, Presented at the 7th TLEP-FCCee workshop, 19-21 June 2014, CERN, URL
  http://indico.cern.ch/event/313708/.

\bibitem{Seidel:2013sqa}
K.~Seidel, F.~Simon, M.~Tesar, S.~Poss, {Top quark mass measurements at and
  above threshold at CLIC}, Eur.Phys.J. C73 (2013) 2530.
\newblock \href {http://arxiv.org/abs/1303.3758} {\path{arXiv:1303.3758}},
  \href {http://dx.doi.org/10.1140/epjc/s10052-013-2530-7}
  {\path{doi:10.1140/epjc/s10052-013-2530-7}}.

\bibitem{Hoang:1999zc}
A.~Hoang, T.~Teubner, {Top quark pair production close to threshold: Top mass,
  width and momentum distribution}, Phys.Rev. D60 (1999) 114027.
\newblock \href {http://arxiv.org/abs/hep-ph/9904468}
  {\path{arXiv:hep-ph/9904468}}, \href
  {http://dx.doi.org/10.1103/PhysRevD.60.114027}
  {\path{doi:10.1103/PhysRevD.60.114027}}.

\bibitem{Hoang:1998xf}
A.~Hoang, T.~Teubner, {Top quark pair production at threshold: Complete
  next-to-next-to-leading order relativistic corrections}, Phys.Rev. D58 (1998)
  114023.
\newblock \href {http://arxiv.org/abs/hep-ph/9801397}
  {\path{arXiv:hep-ph/9801397}}, \href
  {http://dx.doi.org/10.1103/PhysRevD.58.114023}
  {\path{doi:10.1103/PhysRevD.58.114023}}.

\bibitem{Khanpour:2014xla}
H.~Khanpour, S.~Khatibi, M.~K. Yanehsari, M.~M. Najafabadi, {Single top quark
  production as a probe of anomalous $tq\gamma$ and $tqZ$ couplings at the
  FCC-ee}\href {http://arxiv.org/abs/1408.2090} {\path{arXiv:1408.2090}}.

\bibitem{Ellis}
J. Ellis, Presented at the 7th TLEP-FCCee workshop, 19-21 June 2014, CERN, URL
  http://indico.cern.ch/event/313708/.

\end{thebibliography}







\end{document}